\documentclass[aps,english,twocolumn,showpacs]{revtex4}
\usepackage{mathrsfs}
\usepackage{graphicx}
\usepackage{amsmath, amssymb}
\usepackage{babel}
\usepackage{bm}

\begin{document}

\title{Inverted duality of Hubbard model and an equation for Green's functions}
\preprint{1}

\author{Xiao-Yong Feng}
\email[fxyong@hznu.edu.cn]{}
\affiliation{School of Physics, Hangzhou Normal University, Hangzhou 310036, China}
%\keywords{Hubbard model, duality, doublon}

\begin{abstract}
The Hubbard model, a cornerstone in the field of condensed matter physics, serves as a fundamental framework for investigating the behavior of strongly correlated electron systems. This work presents a novel perspective on the model, uncovering its inherent inverted duality which has profound implications for our comprehension of this complex system. Taking advantage of this special mathematical property, we formulated an equation which directly links the electron Green's function to doublon Green's function. Applied to the triangular lattice Hubbard model, our approach yields results consistent with those from state-of-the-art numerical simulations. Our findings open a route to probing electron-correlation dynamics and phase transitions in the Hubbard model, potentially yielding fresh insight into its emergent many-body physics.
\end{abstract}

\maketitle
\section{introduction}
The Hubbard model holds a position in condensed matter physics akin to the Ising model's role in statistical physics\cite{Hubbard1,Hubbard2,Hubbard3,Hubbard4}. Despite its deceptively simple form, it possesses an extremely rich physical content. It is used to explain phenomena such as Mott insulator behavior\cite{Mott1,Mott2}, itinerant ferromagnetism\cite{FM1,FM2,FM3}, antiferromagnetism\cite{AFM1,AFM2,AFM3}, and high-temperature superconductivity\cite{high_Tc1,high_Tc2,high_Tc3,high_Tc4}. Its construction is so simple that it only includes two terms: the hopping term of electrons between adjacent lattice sites and the on-site Coulomb repulsion term. These two terms are diagonalized in momentum space and real space, respectively. However, their combination makes the analysis of the model very challenging. Analytically, a not very small Coulomb term renders perturbation theory ineffective, while the conventional mean-field theories fail to account for the strong quantum fluctuations exhibited by the model. Numerical treatment is also limited by the negative sign problem and the exponentially growing dimension of the Hilbert space with the number of lattice sites.

To navigate these complexities, symmetry analysis emerges as a vital tool, offering insights into the model's properties prior to its solution\cite{symmetry}. The $U(1)$ symmetry inherent in the Hubbard model guarantees that the electron number for each spin orientation is a good quantum number, resulting in a block-diagonal structure of the Hilbert space with respect to electron number. The Hubbard model exhibits an intrinsic $SU(2)$ spin symmetry, with magnetic states arising from the spontaneous breaking of this symmetry. Its particle-hole symmetry ensures a symmetric phase diagram centered around half-filling. The existence of the pseudospin symmetry leads to new collective modes of the model\cite{Zhang}. Furthermore, the model is endowed with $SO(4)$ symmetry\cite{SO4}, driven by two quantum numbers that are intimately connected to the system's superconducting and magnetic characteristics.

In this work, we will reveal the mathematical structure  what we term the "inverted duality" of the model, and from this derive an equation satisfied by the Green's functions of the model, shedding new light on the model's underlying structure and behavior.
\section{the path-integral representation}
The Hamiltonian of the Hubbard model is given by the following expression
\begin{eqnarray}
\hat{H} = -t\sum_{\langle \mathbf{i}\mathbf{j}\rangle\sigma}\hat{c}^{\dagger}_{\mathbf{i}\sigma}\hat{c}_{\mathbf{j}\sigma}-\mu\sum_{\mathbf{i}\sigma}\hat{n}_{\mathbf{i}\sigma}+U\sum_{\mathbf{i}}\hat{n}_{\mathbf{i}\uparrow}\hat{n}_{\mathbf{i}\downarrow},
\end{eqnarray}
here $\langle \mathbf{i}\mathbf{j}\rangle$ denotes a pair of nearest neighbor sites, $\hat{c}^{\dagger}_{\mathbf{i}\sigma}$ creates an electron with spin $\sigma
(= \uparrow or \downarrow)$ at lattice site $\mathbf{i}$, $\hat{n}_{\mathbf{i}\sigma}=\hat{c}^{\dagger}_{\mathbf{i}\sigma}\hat{c}_{\mathbf{i}\sigma}$ is the corresponding occupation number operator and $\mu$ is the chemical potential which is introduced here to control the electron filling.

The partition function for the system is give by
$Z = tr e^{-\beta\hat{H}}$ where $\beta=\frac{1}{k_BT}$ with $k_B$ being the Boltzmann constant. In the path integral representation\cite{path}, the partition function has the form
\begin{equation}\label{phxi}
Z = \int D(\xi^*,\xi)e^{-S(\xi^*,\xi)}
\end{equation}
in which the integral variable $\xi^*$ and $\xi$ are Grassmann numbers,  $D(\xi^*,\xi)= \lim\limits_{M\rightarrow\infty}\prod\limits_{m=1}^M\prod\limits_{\mathbf{i}\sigma}d\xi^*_{\mathbf{i}\sigma,m}d\xi_{\mathbf{i}\sigma,m}$, and the action
\begin{eqnarray}
S=\sum_{\mathbf{i}m\sigma}(n_{\mathbf{i}\sigma,m}-\xi^*_{\mathbf{i}\sigma,m}\xi_{\mathbf{i}\sigma,m-1})+\sum_{m}\epsilon H(\xi^*_{m},\xi_{m-1})
\end{eqnarray}
with $n_{\mathbf{i}\sigma,m}=\xi^*_{\mathbf{i}\sigma,m}\xi_{\mathbf{i}\sigma,m}$, $\epsilon = \beta/M$ and $H$ being the classical Hamiltonian.

After performing the Fourier transformation
\begin{eqnarray}
\xi_{\mathbf{i}\sigma,m}&=&\frac{1}{\sqrt{N\beta}}\sum_{k}\xi_{k\sigma}e^{-i\omega_n\tau_m+i\mathbf{k}\cdot \mathbf{i}},\\
\xi_{k\sigma}&=&\frac{\epsilon}{\sqrt{N\beta}}\sum_{\mathbf{i}m}\xi_{\mathbf{i}\sigma,m}e^{i\omega_n\tau_m-i\mathbf{k}\cdot \mathbf{i}}
\end{eqnarray}
for any Bravais lattice, we have
\begin{eqnarray}\label{S}
S=-\sum_{k\sigma}(i\omega_n-\varepsilon_{k})\xi^*_{k\sigma}\xi_{k\sigma}+\epsilon U\sum_{\mathbf{i}m}\prod_\sigma\xi^*_{\mathbf{i}\sigma,m}\xi_{\mathbf{i}\sigma,m-1}
\end{eqnarray}
where $\omega_n = (2n-1)\pi/\beta$  ($n$ is an integer ranging from $1$ to $M$.) are the fermionic  Matsubara frequencies, $\tau_m=m\epsilon$, $k=(i\omega_n,\mathbf{k})$,
$\varepsilon_{k}=(\varepsilon_{\mathbf{k}}-\mu)\exp{(i\omega_n\epsilon)}$,  $\varepsilon_{\mathbf{k}}$ is the energy band of free electrons and $\exp{(i\omega_n\epsilon)}$ is kept here for correct causality in the terminology of field theory.

\section{the decoupling of the quadratic term}
The quadratic term in the action can be decoupled
 by employing the following identity\cite{identity},%Adopting the way of the dual fermion approach\cite{dual}, we decouple the quadratic term in the action with the following identity,
\begin{eqnarray}
e^{\alpha\xi^*\xi} = \alpha\int d\eta^*d\eta
e^{-\alpha^{-1}\eta^*\eta+\eta^*\xi+\xi^*\eta}
\end{eqnarray}
in which $\eta^*$ and $\eta$ are Grassmann numbers and $\alpha$ is generally a complex number. It is worth noting this method of handling the quadratic term has previously been utilized in the dual fermion approach \cite{dual1,dual2}.
Then, the partition function becomes
\begin{eqnarray}\label{pfS}
Z&=&\int \prod_{k\sigma}(i\omega_n-\varepsilon_{k})d\eta^*_{k\sigma}d\eta_{k\sigma}e^{-\sum\limits_{k\sigma}(i\omega_n-\varepsilon_{k})^{-1}\eta_{k\sigma}^*\eta_{k\sigma}}\nonumber\\&&\times\int D(\xi^*,\xi)\prod_{\mathbf{i}}e^{-A_\mathbf{i}}
\end{eqnarray}
in which the action for the $\xi$ field is decoupled with respect to the lattice site and the action for a single lattice site takes the form
\begin{eqnarray}
A_\mathbf{i} &=&-\sum_{m\sigma}\epsilon(\eta^*_{\mathbf{i}\sigma,m}\xi_{\mathbf{i}\sigma,m}+\xi_{\mathbf{i}\sigma,m}^*\eta_{\mathbf{i}\sigma,m})\nonumber\\&&+\sum_{m}\epsilon U \xi^*_{\mathbf{i}\uparrow,m}\xi_{\mathbf{i}\uparrow,m-1}\xi^*_{\mathbf{i}\downarrow,m}\xi_{\mathbf{i}\downarrow,m-1}.
\end{eqnarray}

Upon series expansion, we have
\begin{eqnarray}\label{expand}
e^{-A_\mathbf{i}} &=&\prod_{m\sigma}\left(1+ \epsilon\eta^{*}_{\mathbf{i}\sigma,m}\xi_{\mathbf{i}\sigma,m}\right)\left(1+\epsilon\xi_{\mathbf{i}\sigma,m}^*\eta_{\mathbf{i}\sigma,m}\right)\nonumber\\&&\prod_{m}\left(1-\epsilon U \xi^*_{\mathbf{i}\uparrow,m}\xi_{\mathbf{i}\uparrow,m-1}\xi^*_{\mathbf{i}\downarrow,m}\xi_{\mathbf{i}\downarrow,m-1}\right).
\end{eqnarray}

\section{the diagram technique}

We manage to integrate out the $\xi$ field from partition function given by expression (\ref{pfS}).
According to the integration rules of Grassmann numbers, the terms contributing to the integral of $\xi$ field are ones constructed by multiplying all the $\xi^*_{\mathbf{i}\sigma,m}$ and $\xi_{\mathbf{i}\sigma,m}$ once and only once for each lattice site $\mathbf{i}$. In order to intuitively identify the terms that contribute, we use a diagram technique.

\begin{figure}[ht]
\includegraphics [width=5cm]{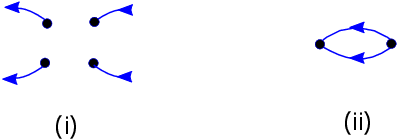}
 \caption{The building blocks.} \label{fig1}
\end{figure}

The building blocks of the terms that contribute in expression (\ref{expand}) are depicted in Figure \ref{fig1}.  The dot represents the position of imaginary-time slice and the line with arrow coming from or pointing to the slice $m$ represents $\xi_{\mathbf{i}\sigma,m}$  and $\xi^*_{\mathbf{i}\sigma,m}$ respectively. The lines above the dots are for the spin up component and the ones below the dots are for the spin down component. Therefore the building block of type (ii) represents $\xi^*_{\mathbf{i}\uparrow,m}\xi_{\mathbf{i}\uparrow,m-1}\xi^*_{\mathbf{i}\downarrow,m}\xi_{\mathbf{i}\downarrow,m-1}$.

 When all the dots are connected by paired incoming and outgoing lines for each spin component - meaning the term takes a form as $\prod_{m\sigma}\xi^*_{\mathbf{i}\sigma,m}\xi_{\mathbf{i}\sigma,m}$, the corresponding Grassmann integral becomes non-vanishing. We call such a configuration the contribution diagram. The contribution of a contribution diagram to the integral is given by the product of the coefficients carried by each building block participating in the contribution diagram.

 \section{the inverted duality}

\begin{figure}[ht]
\includegraphics [width=8cm]{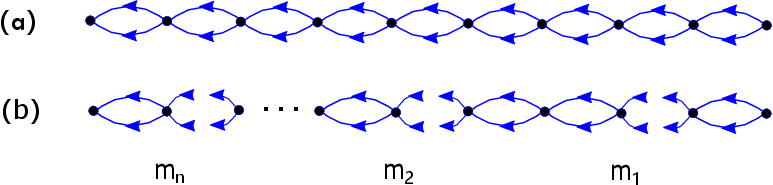}
 \caption{The contribution diagrams for the integral of $\xi$ field.} \label{fig2}
\end{figure}

All the contribution diagrams for the integral of $\xi$ field in the expression (\ref{pfS}) are illustrated in Figure \ref{fig2}.  Figure \ref{fig2} (a) shows the contribution diagram constructed solely by building block of type (ii) while Figure \ref{fig2} (b) shows the ones in the absence of $n$ ($n\geq1$) building blocks of type (ii) which are replaced by building blocks of type (i). The corresponding contributions  are \begin{eqnarray}
(a)&:&(-\epsilon U)^M, \\
(b)&:&(-\epsilon U)^{M-n}\sum_{m_n>\cdots>m_2>m_1}\prod_{l=1}^n\epsilon^4\prod_{\sigma}\eta^{*}_{\mathbf{i}\sigma,m_l-1}\eta_{\mathbf{i}\sigma,m_l}\nonumber\\
&=&(-\epsilon U)^M\frac{1}{n!}\left(-\frac{\epsilon^3}{ U}\sum_m\prod_{\sigma}\eta^{*}_{\mathbf{i}\sigma,m-1}\eta_{\mathbf{i}\sigma,m}\right)^n.
\end{eqnarray}

Gathering the above terms gives
\begin{eqnarray}
(-\epsilon U)^Me^{-\frac{\epsilon^3}{ U}\sum\limits_m\eta^{*}_{\mathbf{i}\uparrow,m-1}\eta_{\mathbf{i}\uparrow,m}\eta^{*}_{\mathbf{i}\downarrow,m-1}\eta_{\mathbf{i}\downarrow,m}}.
\end{eqnarray}

Therefore, after integrating out the $\xi$ field, the partition function becomes
\begin{eqnarray}
Z=(- \epsilon U)^{MN}\prod_{k\sigma}(i\omega_n-\varepsilon_{k})\int D'(\eta^*,\eta)e^{-S'(\eta^*,\eta)}
\end{eqnarray}
in which $D'(\eta^*,\eta)=\prod\limits_{k\sigma}d\eta^*_{k\sigma}d\eta_{k\sigma}$ and the action  is
\begin{eqnarray}
S'=\sum_{k\sigma}\frac{1}{i\omega_n-\varepsilon_{k}}\eta_{k\sigma}^*\eta_{k\sigma}+\frac{\epsilon ^3}{ U}\sum_{\mathbf{i}m}\prod_{\sigma}\eta^{*}_{\mathbf{i}\sigma,m-1}\eta_{\mathbf{i}\sigma,m}.
\end{eqnarray}
It is easy to show that after a particle-hole transformation $\eta(\eta^*)\rightarrow\eta^*(\eta)$, comparing with expression (\ref{S}), the quantity before the quadratic term is inverted and so does the Coulomb repulsion $U$. Thus, we find that there is an inverted duality between the actions of  $\xi$ and $\eta$ fields.

Such a duality becomes more pronounced when we perform the same operation on the action of $\eta$ field as on the one of $\xi$ field. To this end, we introduce another Grassmann field $\zeta$ and decouple the quadratic term in action $S'$ as below
\begin{eqnarray}
e^{-\alpha^{-1}\eta^*\eta} = -\frac{1}{\alpha}\int d\zeta^*d\zeta
e^{\alpha\zeta^*\zeta+\zeta^*\eta+\eta^*\zeta}.
\end{eqnarray}
And the partition function becomes
\begin{eqnarray}\label{pfSp}
Z &=&(-\epsilon U)^{MN}\int \prod_{k\sigma}d\zeta^*_{k\sigma}d\zeta_{k\sigma}e^{\sum\limits_{k\sigma}(i\omega_n-\varepsilon_{k})\zeta_{k\sigma}^*\zeta_{k\sigma}}\nonumber\\&&\times\epsilon ^{-2MN}\int\prod_{\mathbf{i}m\sigma}d\eta^*_{\mathbf{i}\sigma,m}d\eta_{\mathbf{i}\sigma,m}\prod_{\mathbf{i}}e^{-A'_\mathbf{i}}
\end{eqnarray}
where $\epsilon ^{-2MN}$ is the Jacobian determinant of the Fourier transformation from $\eta_{k\sigma}$ to $\eta_{\mathbf{i}\sigma,m}$ and
\begin{eqnarray}
A'_\mathbf{i} & = & -\sum_{\mathbf{i}m\sigma}\epsilon(\zeta^{*}_{\mathbf{i}\sigma,m}\eta_{\mathbf{i}\sigma,m}+\eta_{\mathbf{i}\sigma,m}^*\zeta_{\mathbf{i}\sigma,m})\nonumber\\&& +\sum_{\mathbf{i}m} \frac{\epsilon ^3}{ U}\eta^{*}_{\mathbf{i}\uparrow,m-1}\eta_{\mathbf{i}\uparrow,m}\eta^{*}_{\mathbf{i}\downarrow,m-1}\eta_{\mathbf{i}\downarrow,m}.
\end{eqnarray}

The series expansion gives
\begin{eqnarray}\label{expand2}
e^{-A'_\mathbf{i}} &=&\prod_{m\sigma}\left(1+ \epsilon\zeta^{*}_{\mathbf{i}\sigma,m}\eta_{\mathbf{i}\sigma,m}\right)\left(1+\epsilon\eta_{\mathbf{i}\sigma,m}^*\zeta_{\mathbf{i}\sigma,m}\right)\nonumber\\&&\prod_{m}\left(1-\frac{\epsilon ^3}{ U} \eta^*_{\mathbf{i}\uparrow,m-1}\eta_{\mathbf{i}\uparrow,m}\eta^*_{\mathbf{i}\downarrow,m-1}\eta_{\mathbf{i}\downarrow,m}\right).
\end{eqnarray}

It is worth noting that due to the reversed imaginary-time order in $A'_\mathbf{i}$, the arrows in the contribution diagrams for the integral of $\eta$ field are just reversed comparing with Figure \ref{fig2}. The summation of their contributions gives
\begin{eqnarray}
\left(-\frac{\epsilon^3}{U}\right)^Me^{-\epsilon U\sum\limits_m\zeta^{*}_{\mathbf{i}\uparrow,m}\zeta_{\mathbf{i}\uparrow,m-1}\zeta^{*}_{\mathbf{i}\downarrow,m}\zeta_{\mathbf{i}\downarrow,m-1}}.
\end{eqnarray}

Then, by integrating  $\eta$ field, we have
\begin{equation}\label{pfzeta}
Z = \int D(\zeta^*,\zeta)e^{-S''(\zeta^*,\zeta)}
\end{equation}
with $D(\zeta^*,\zeta)= \lim\limits_{M\rightarrow\infty}\prod\limits_{m=1}^M\prod\limits_{\mathbf{i}\sigma}d\zeta^*_{\mathbf{i}\sigma,m}d\zeta_{\mathbf{i}\sigma,m}$ (The Jacobian determinant of the Fourier transformation from $\zeta_{k\sigma}$ to $\zeta_{\mathbf{i}\sigma,m}$ has already been taken into account.) and
\begin{eqnarray}
S''=-\sum_{k\sigma}(i\omega_n-\varepsilon_{k})\zeta^*_{k\sigma}\zeta_{k\sigma}+\epsilon U\sum_{\mathbf{i}m}\prod_\sigma\zeta^*_{\mathbf{i}\sigma,m}\zeta_{\mathbf{i}\sigma,m-1}.
\end{eqnarray}

It is found that the partition function constructed by the integration of  $\zeta$ field shares exactly the same structure as the one constructed by $\xi$ field.  So far, we have fully demonstrated the inverted duality of the Hubbard model and reveal the $Z_2$ symmetry of the dual transformation operation.

\section{An equation for Green's functions}

Next, we utilize the duality of the model to derive an equation about the Green's functions.

By adding an external source term $\sum\limits_{k\sigma}(\eta^*_{k\sigma}J_{\xi,k\sigma}+J^*_{\xi,k\sigma}\eta_{k\sigma})$ to action of the partition function given by expression (\ref{pfS}), we have
\begin{eqnarray}\label{d1}
\langle\eta^*_{k\sigma}\eta_{k\sigma}\rangle = \left.-\frac{1}{Z[J_{\xi}]}\frac{\partial^2 Z[J_{\xi}]}{\partial J_{\xi,k\sigma}\partial J^{*}_{\xi,k\sigma}}\right|_{J_{\xi}=0}
\end{eqnarray}
where $Z[J_{\xi}]$ is the partition function with the external source,  $\langle \cdots\rangle $ denotes the ensemble average and $\langle\eta^*_{k\sigma}\eta_{k\sigma}\rangle$ is a Green's function of $\eta$ field.

Integrating out $\eta$ field from $Z[J_{\xi}]$  leads to
\begin{eqnarray}
Z[J_{\xi}] = \int D(\xi^*,\xi)e^{-S[J_{\xi}]}
\end{eqnarray}
with
\begin{eqnarray}
S[J_{\xi}]&=&-\sum_{k\sigma}(i\omega_n-\varepsilon_{k})(\xi^*_{k\sigma}+ J^{*}_{\xi,k\sigma})(\xi_{k\sigma}+J_{\xi,k\sigma})\nonumber\\&&+\epsilon U\sum_{\mathbf{i}m}\xi^*_{\mathbf{i}\uparrow,m}\xi_{\mathbf{i}\uparrow,m-1}\xi^*_{\mathbf{i}\downarrow,m}\xi_{\mathbf{i}\downarrow,m-1}.
\end{eqnarray}
Therefore, the second-order partial derivatives of  $Z[J_{\xi}]$ in expression (\ref{d1}) leads to
\begin{eqnarray}\label{eta1}
\langle\eta^*_{k\sigma}\eta_{k\sigma}\rangle =(i\omega_n-\varepsilon_{k})^2\langle\xi^*_{k\sigma}\xi_{k\sigma}\rangle -(i\omega_n-\varepsilon_{k}).
\end{eqnarray}

On the other hand, by adding an external source term $\sum\limits_{k\sigma}(\eta^*_{k\sigma}J_{\zeta,k\sigma}+J^*_{\zeta,k\sigma}\eta_{k\sigma})$ to action of the partition function given by expression (\ref{pfSp}) (At this time, the partition function is denoted as  $Z[J_{\zeta}]$.), the Green's function of $\eta$ field can be expressed as
\begin{eqnarray}\label{d2}
\langle\eta^*_{k\sigma}\eta_{k\sigma}\rangle = \left.-\frac{1}{Z[J_{\zeta}]}\frac{\partial^2 Z[J_{\zeta}]}{\partial J_{\zeta,k\sigma}\partial J^{*}_{\zeta,k\sigma}}\right|_{J_{\zeta}=0}.
\end{eqnarray}

The integration with respect to $\eta$ field leads to
\begin{equation}
Z[J_{\zeta}] = \int D(\zeta^*,\zeta)e^{-S[J_{\zeta}]}
\end{equation}
with
\begin{eqnarray}
S[J_{\zeta}]&=&-\sum_{k\sigma}(i\omega_n-\varepsilon_{k})\zeta^*_{k\sigma}\zeta_{k\sigma}\nonumber\\&&+\epsilon U\sum_{\mathbf{i}m}\zeta^{(J)*}_{\mathbf{i}\uparrow,m}\zeta^{(J)}_{\mathbf{i}\uparrow,m-1}\zeta^{(J)*}_{\mathbf{i}\downarrow,m}\zeta^{(J)}_{\mathbf{i}\downarrow,m-1}
\end{eqnarray}
where $\zeta^{(J)}=\zeta+J_{\zeta}$.

The Fourier transformation of the $U$ term on the above action is
\begin{eqnarray}
&&\epsilon U\sum_{\mathbf{i}m}\zeta^{(J)*}_{\mathbf{i}\uparrow,m}\zeta^{(J)}_{\mathbf{i}\uparrow,m-1}\zeta^{(J)*}_{\mathbf{i}\downarrow,m}\zeta^{(J)}_{\mathbf{i}\downarrow,m-1}\nonumber\\
&=&\frac{U}{N\beta}\sum_{kk'q'}\zeta^{(J)*}_{k\uparrow}\zeta^{(J)}_{k+q'\uparrow}\zeta^{(J)*}_{k'\downarrow}\zeta^{(J)}_{k'-q'\downarrow}
\end{eqnarray}
where $q'=(i\Omega'_n,\mathbf{q})$ with $\Omega'_n=2n\pi/\beta$ being bosonic Matsubara frequencies. Then the second-order partial derivatives of $Z[J_{\zeta}]$ in the expression (\ref{d2}) leads to
\begin{eqnarray}\label{eta2}
\langle\eta^*_{k\sigma}\eta_{k\sigma}\rangle= \frac{U}{N\beta}\sum_{k'}\langle\zeta^*_{k'\bar{\sigma}}\zeta_{k'\bar{\sigma}}\rangle+\left( \frac{U}{N\beta}\right)^2F_{k\sigma}
\end{eqnarray}
where $\bar{\sigma}$ denotes the spin opposite to $\sigma$ and
\begin{eqnarray}
F_{k\sigma}=\sum_{k'k''q'q''}\langle\zeta^{*}_{k-q''\sigma}\zeta^{*}_{k''\bar{\sigma}}\zeta_{k''-q''\bar{\sigma}}\zeta_{k+q'\sigma}\zeta^{*}_{k'\bar{\sigma}}\zeta_{k'-q'\bar{\sigma}}\rangle.
\end{eqnarray}

Since the partition functions (\ref{phxi}) and (\ref{pfzeta}) own identical mathematical structure,  the Green's functions of $\xi$ field should be equal to the ones of $\zeta$ field. Applying this relationship and combining expressions (\ref{eta1}) and (\ref{eta2}), we finally obtain that
\begin{eqnarray}\label{equation}
\langle\xi^*_{k\sigma}\xi_{k\sigma}\rangle =\frac{1}{i\omega_n-\varepsilon_{k}}+\frac{Un_{\bar{\sigma}}}{(i\omega_n-\varepsilon_{k})^2}
+\frac{\left( \frac{U}{N\beta}\right)^2G_{k\sigma}}{(i\omega_n-\varepsilon_{k})^2}
\end{eqnarray}
where $n_{\bar{\sigma}}= \frac{1}{N\beta}\sum\limits_{k'}\langle\xi^*_{k'\bar{\sigma}}\xi_{k'\bar{\sigma}}\rangle$ as the density of electrons with $\bar{\sigma}$ spin is a conserved quantity of the system and
$
G_{k\sigma}=\sum\limits_{k'k''q'q''}\langle\xi^{*}_{k-q''\sigma}\xi^{*}_{k''\bar{\sigma}}\xi_{k''-q''\bar{\sigma}}\xi_{k+q'\sigma}\xi^{*}_{k'\bar{\sigma}}\xi_{k'-q'\bar{\sigma}}\rangle
$
is the sum of a set of Matsubara Green's functions \cite{Matsubara} and it can be reexpressed as

\begin{eqnarray}\label{G}
G_{k\sigma}&=&\beta^2\sum_{\mathbf{k}'\mathbf{k}''\mathbf{q}'\mathbf{q}''}\ll\hat{A}_{\mathbf{k}\mathbf{k}'\mathbf{q}'\sigma}|\hat{B}_{\mathbf{k}\mathbf{k}''\mathbf{q}''\sigma}\gg_{i\omega_n}\nonumber\\
&=&(N\beta)^2\ll \hat{d}_{\textbf{k}\sigma}|\hat{d}^{\dag}_{\textbf{k}\sigma}\gg_{i\omega_n}
\end{eqnarray}
among which $\hat{A}_{\mathbf{k}\mathbf{k}'\mathbf{q}'\sigma}=\hat{c}_{\mathbf{k}+\mathbf{q}'\sigma}\hat{c}^{\dag}_{\mathbf{k}'\bar{\sigma}}\hat{c}_{\mathbf{k}'-\mathbf{q}'\bar{\sigma}}$ , $\hat{B}_{\mathbf{k}\mathbf{k}''\mathbf{q}''\sigma}=\hat{c}^{\dag}_{\mathbf{k}-\mathbf{q}''\sigma}\hat{c}^{\dag}_{\mathbf{k}''\bar{\sigma}}\hat{c}_{\mathbf{k}''-\mathbf{q}''\bar{\sigma}}$  and $\hat{d}_{\textbf{k}\sigma}$ is the Fourier transformation of the doublon  annihilation operator $\hat{d}_{\textbf{i}\sigma}\equiv\hat{c}_{\textbf{i}\sigma}\hat{n}_{\textbf{i}\bar{\sigma}}$.

The expression (\ref{equation}) is the equation satisfied by the Green's functions of the Hubbard model. This equation indicates that the electron Green's function is directly related to the doublon Green's function. Since the electron annihilation operator can be decomposed as $\hat{c}_{\sigma} = \hat{h}^{\dag}_{\sigma}+ \hat{d}_{\sigma} $ with $\hat{h}^{\dag}_{\sigma} = \hat{c}_{\sigma} (1 - \hat{n}_{\bar{\sigma}})$ being  holon operator
and $\hat{d}_{\sigma} = \hat{c}_{\sigma} \hat{n}_{\bar{\sigma}}$ being doublon operator respectively\cite{doublon}, the single-particle electron Green's function
necessarily comprises the sum of the holon, doublon and cross Green's functions. Our derived equation provides additional connection between them.  This is the key result of this work.

\section{An application to triangular lattice Hubbard model}
We calculate $G_{k\sigma}$ with the method of motion equation\cite{Green1}. The equation of motion for the Matsubara Green's functions is written as \cite{Green2}
\begin{eqnarray}\label{Green}
i\omega_n\ll\hat{A}|\hat{B}\gg_{i\omega_n}= \langle[\hat{A},\hat{B}]_{+}\rangle+\ll[\hat{A},\hat{H}]|\hat{B}\gg_{i\omega_n}.
\end{eqnarray}
For the sake of brevity, the subscripts of operators $\hat{A}_{\mathbf{k}\mathbf{k}'\mathbf{q}'\sigma}$ and $\hat{B}_{\mathbf{k}\mathbf{k}''\mathbf{q}''\sigma}$  have been omitted here and in the subsequent discussions.

The terms involved in the equation of motion are calculated as follows
\begin{eqnarray}
&&\sum_{\mathbf{k}'\mathbf{k}''\mathbf{q}'\mathbf{q}''}\langle[\hat{A},\hat{B}]_{+}\rangle=N\sum_{\mathbf{i}}\langle\hat{n}_{\mathbf{i}\bar{\sigma}}\rangle=N^2n_{\bar{\sigma}} ,\\
&&\sum_{\mathbf{k}'\mathbf{q}'}[\hat{A},\hat{H}]=(U-\mu)\sum_{\mathbf{k}'\mathbf{q}'}\hat{A}+\sum_{\mathbf{k}'\mathbf{q}'}E_{\mathbf{k}\mathbf{k}'\mathbf{q}'}\hat{A}
\end{eqnarray}
in which $E_{\mathbf{k}\mathbf{k}'\mathbf{q}'}=\varepsilon_{\mathbf{k}+\mathbf{q}'}+\varepsilon_{\mathbf{k}'-\mathbf{q}'}-\varepsilon_{\mathbf{k}'}$.

The summation of motion equation with respect to $\mathbf{k}'$, $\mathbf{q}'$, $\mathbf{k}''$ and $\mathbf{q}''$ gives
\begin{eqnarray}\label{E}
\sum_{\mathbf{k}'\mathbf{k}''\mathbf{q}'\mathbf{q}''}(i\omega_n-E_{\mathbf{k}\mathbf{k}'\mathbf{q}'}+\mu-U)\ll\hat{A}|\hat{B}\gg_{i\omega_n}=N^2n_{\bar{\sigma}}
\end{eqnarray}

Under the condition of no nesting at the Fermi surface (The nesting of the Fermi surface will introduce scattering processes near $\mathbf{q}'=\pi$ ), considering only the scattering processes that occur between states with a very small momentum difference, i.e. $\mathbf{q}'\approx0$,  we have $E_{\mathbf{k}\mathbf{k}'\mathbf{q}'}\approx\varepsilon_{\mathbf{k}}$ and the expression (\ref{E}) gives
\begin{eqnarray}
\sum_{\mathbf{k}'\mathbf{k}''\mathbf{q}'\mathbf{q}''}\ll\hat{A}|\hat{B}\gg_{i\omega_n}\approx\frac{N^2n_{\bar{\sigma}}}{i\omega_n - (\varepsilon_{\mathbf{k}} - \mu + U)}.
\end{eqnarray}
Therefore, according to expression (\ref{G}),
\begin{eqnarray}
G_{k\sigma} = \frac{(N\beta)^2 n_{\bar{\sigma}}}{i\omega_n-\varepsilon_{k} - U}.
\end{eqnarray}

By substituting above expression into the equation (\ref{equation}), we arrive at
\begin{eqnarray}\label{spectrum}
\langle\xi^*_{k\sigma}\xi_{k\sigma}\rangle =\frac{1-n_{\bar{\sigma}}}{i\omega_n-\varepsilon_{k}}+\frac{n_{\bar{\sigma}}}{i\omega_n-\varepsilon_{k}-U}.
\end{eqnarray}

According to this expression, the spectrum of the Hubbard model is split into upper and lower Hubbard bands $\varepsilon_{k}+U$ and $\varepsilon_{k}$ with spectral weights $n_{\bar{\sigma}}$ and $1-n_{\bar{\sigma}}$, respectively . Once the Coulomb interaction U becomes larger than the bandwidth W of the free hopping electron, a Mott gap emerges between the upper and lower Hubbard bands , and the system enters an insulating state. That is to say, the Mott transition occurs at $U_c/W=1$.

The analysis above has provided a concise formula for the Mott transition. This result is beyond the Hubbard I approximation \cite{Hubbard1} which always leads to a finite gap between the Hubbard bands, regardless of the lattice type. And it is also  different from the Hubbard III treatment \cite{Hubbard2} in which the scattering correction and the resonance broadening correction are made by truncating the motion equations at higher orders. Here, we directly deal with higher-order Green's functions without invoking any explicit mean-field-type truncation: the scattering correction terms appear in form of  $\ll\hat{A}|\hat{B}\gg_{i\omega_n}$  requiring no truncation and the resonance broadening correction terms just disappear when we taking the approximation of small-momentum scattering which is appropriate as long as scattering in Fermionic systems mainly occurs near the surface of the Fermi sea. It must be recognized that for an interacting system, the decay of quasiparticles is always present, which cannot be given by the approximation of small-momentum scattering
, but its effect is merely to broaden energy levels and does not fundamentally alter the physical picture of the Mott transition.

It is worth mentioning that the intricacies of the Mott transition are particularly nuanced\cite{review1}. There is a charge gap for any nonvanishing U in the Hubbard model on a square lattice\cite{Uc01,Uc02}. The formation of the gap is believed to be due to the perfect nesting of the Fermi surface\cite{Slater,Uc1}. Such a mechanism has not yet been considered here.  In the Hubbard model on a hexagonal lattice, there indeed exists a finite critical value $U_{c}$ that separates the metallic phase from the Mott insulating phase\cite{Uc2}. However, our formulae do not readily extend to hexagonal lattice since it does not belong to Bravais lattice.

One of the lattices suitable for formula (\ref{spectrum}) is the triangular lattice. In an authoritative study combining multiple powerful numerical methods\cite{Mott3}, the spectral function $A_c(\omega=0)$ of the triangular lattice Hubbard model calculated with cellular dynamical mean-field theory exhibits distinct temperature-dependent behaviors on either side of $U_c/W=1$ (here $W=9t$), where the slope of the kinetic energy also demonstrates a discontinuity. While the calculation of static charge structure factor with the minimally entangled
thermal typical state method at low temperature gives $U_c/W\approx0.97$. These indicate that the Mott transition in the triangular lattice Hubbard model occurs at $U_c/W\approx1$. The result we obtained, $U_c/W=1$, is in excellent agreement with the aforementioned numerical findings. That is to say, for the triangular lattice, in the absence of Fermi surface nesting, the small-momentum scattering approximation is applicable, and our results provide a quantitatively correct Mott transition point. Of course, more advanced methods are required to calculate the doublon Green's function in order to investigate other properties in the Hubbard model\cite{Roth,Lee,Haurie,Pangburn,Banerjee}, which we've reserved for future exploration.
%Despite the existence of a spin liquid phase prior to the Mott transition remains a matter of debate\cite{Uc3,Uc4}.

\section{summary}
In summary, we have demonstrated the existence of an inverted duality in the Hubbard model and used it to derive an equation which directly links the electron Green's function to the doublon Green's function.  A preliminary analysis provided us a concise formula for the Mott transition and its application to the Hubbard model on a triangular lattice is very successful. This advancement not only deepens our understanding of the Hubbard model but also offers an alternative approach to dealing with the electron Green's function in the Hubbard model.

\section* {Acknowledgments}
The author appreciates useful discussions with Prof. J. Dai and encouragement from Prof. T. Xiang's group.

\end{document}